\documentclass[aps,prb,reprint]{revtex4-1}
\usepackage{amsmath,amssymb,bm}
\usepackage{graphicx}
\usepackage[caption=false]{subfig}
\usepackage[bookmarks=false,breaklinks=true]{hyperref}
\usepackage{siunitx}
\usepackage{xcolor}

\hypersetup{
  pdfstartview = FitH,
  pdftoolbar   = false,
  pdfmenubar   = true,
  colorlinks   = true,
  linkcolor    = blue!50!black,
  urlcolor     = blue!50!black,
  citecolor    = blue!50!black
}

\begin{document}

\title{Many-body dynamics of chemically propelled nanomotors}
\author{Peter~H.~Colberg}
\email{pcolberg@chem.utoronto.ca}
\author{Raymond~Kapral}
\email{rkapral@chem.utoronto.ca}
\affiliation{Chemical Physics Theory Group, Department of Chemistry,
University of Toronto, Toronto, Ontario M5S 3H6, Canada}

\date{\today}

\begin{abstract}
The collective behavior of chemically propelled sphere-dimer motors made from
linked catalytic and noncatalytic spheres in a quasi-two-dimensional confined
geometry is studied using a coarse-grained microscopic dynamical model.
Chemical reactions at the catalytic spheres that convert fuel to product
generate forces that couple to solvent degrees of freedom as a consequence of
momentum conservation in the microscopic dynamics.
The collective behavior of the many-body system is influenced by direct
intermolecular interactions among the motors, chemotactic effects due to
chemical gradients, hydrodynamic coupling, and thermal noise.
Segregation into high and low density phases and globally homogeneous states
with strong fluctuations are investigated as functions of the motor
characteristics.
Factors contributing to this behavior are discussed in the context of active
Brownian models.
\end{abstract}

\maketitle

\section{Introduction} \label{sec:intro}

Self-organization and pattern formation are processes that occur in systems
under both equilibrium and nonequilibrium conditions.
The principles underlying the dynamics leading to the final structures are well
understood, although in specific applications the evolution equations and the
final inhomogeneous states may be very complicated.%
~\cite{cross1993,hoyle2006,pismen2006,desai2009}
In equilibrium systems, the phase ordering or segregation processes are often
described in terms of order parameter fields, satisfying certain symmetries,
whose evolution is controlled by equations of motion based on free energy
functionals.
In nonequilibrium systems, in general, no such free energy functionals exist,
and the order parameter equations are typically derived from a nonlinear
analysis.
In this case, the variety of possible inhomogeneous states is large and their
classification is more difficult.

Active systems, where the constituent objects either are forced by external
means or move autonomously, are another class of nonequilibrium systems that
self-organize in ways that differ from equilibrium and nonequilibrium systems
with inactive elements.
Systems of this type take many forms including shaken granular materials, a
wide rage of biological systems, and synthetic active media.
Although active systems have been intensively studied only recently, a
considerable literature exists, documented in
reviews~\cite{lauga2009,marchetti2013,elgeti2015,zoettl2016,bechinger2016},
which describes the new phenomena that these systems exhibit.

The models that are used to describe the collective motions of active systems
range from simple particle-based models~\cite{vicsek1995,chate2008,peruani2006}
to models based on continuum order-parameter
equations~\cite{toner1995,simha2002,toner2005}.
Some of the simplest models deal with active Brownian particles that have a
prescribed velocity, interact through hard or soft potentials, and experience
translational and orientational Brownian motion.%
~\cite{fily2012,redner2013,cates2013,stenhammar2013,bialke2013,wysocki2014,fily2014,speck2014,stenhammar2014,takatori2015,speck2015}
Even with these minimal ingredients, the systems separate into high and low
density phases for high enough volume or area fractions, and such dynamics has
been mapped onto continuum order-parameter equations.%
~\cite{stenhammar2013,speck2014}

In many systems where active particles move in a fluid environment, particle
motion induces fluid flows in the surrounding medium that lead to hydrodynamic
coupling among the particles.
Investigations have shown that such hydrodynamic coupling can give rise to
coherent collective motion.%
~\cite{saintillan2012,zoettl2014,oyama2016}

The collective dynamics of self-propelled motors whose motion is governed by
diffusiophoretic mechanisms~\cite{anderson1989,golestanian2005,kapral2013}
is considered in this paper.
Such chemically powered motors operate under nonequilibrium conditions, and
catalytic chemical reactions on the motor generate concentration gradients of
the reactive species that are responsible for motor motion.
The concentration and fluid velocity fields that arise from the activity of all
motors in the system influence the nature of the collective dynamics.
Experimental studies of the collective behavior of chemically powered active
colloidal particles have shown that these systems display clustering, schooling,
and other collective states.%
~\cite{ibele2009,theurkauff2012,palacci2013,buttinoni2013,wang2015}
Langevin models without hydrodynamic interactions have been developed and used
to construct phase diagrams that show the possible types of collective behavior
that such systems can exhibit.%
~\cite{pohl2014,saha2014,pohl2015}

We investigate the collective dynamics of sphere-dimer motors using a
coarse-grained microscopic dynamical method that accounts for coupling through
many-body hydrodynamic interactions, concentration gradients, and direct
potential interactions among motors, as well as thermal fluctuations.
Most studies of the collective dynamics of chemically powered motors are
carried out on spherical Janus particles with catalytic and noncatalytic
hemispherical faces.
The sphere-dimer motors~\cite{rueckner2007,tao2008,valadares2010,yang2014,reigh2015}
we study here have a different shape, are made from linked catalytic and
noncatalytic spheres, and are propelled by a diffusiophoretic mechanism.
Simulations of small numbers of sphere-dimer motors showed that they
self-assemble into transient clusters and display schooling behavior.%
~\cite{thakur2012,kapral2014}
Here we consider systems with thousands of motors and quantitatively
characterize the collective dynamics.
Because our particle-based dynamics describes all interactions and chemical
reactions, and conserves mass, momentum, and energy, it captures the full
many-body structures of the hydrodynamic velocity and concentration fields
without resorting to simplifying approximations, such as the neglect of
correlations in mean field models.
It also accounts for the modification of the propulsion properties of
individual motors due to chemical gradients arising from other
motors in the system.
As a result, we are able to probe aspects of the collective behavior that
are not accessible using other models.

Sec.~\ref{sec:model} contains a description of the coarse-grained microscopic
model for the chemically powered sphere-dimer motor and the solvent in which it
resides.
The properties of a single sphere-dimer motor in solution are given in
Sec.~\ref{sec:single}.
This section also summarizes the continuum description of this motor.
Simulation results on the collective dynamics of large ensembles of
motors are presented and analyzed in Sec.~\ref{sec:collective}.
The results are analyzed further in Sec.~\ref{sec:factors} where connections
with active Brownian particle models are made.
The conclusions of the paper are in Sec.~\ref{sec:conc}.

\section{Microscopic model for collective motor dynamics}\label{sec:model}

The coarse-grained particle-based model for the system includes both the motors
and the multi-component fluid environment in which they move.
The evolution of the entire system is carried out using a hybrid scheme that
combines molecular dynamics (MD) and multiparticle collision dynamics (MPCD).%
~\cite{malevanets1999,*malevanets2000}
The system is contained in a slab whose height is fixed at $L_Z = 30$ and whose
edge lengths $L_X = L_Y = L$ vary as specified below.
The solvent consists of point-like fuel (A) and product (B) molecules with
common mass $m_\text{s} = 1$ and total number density $\varrho_\text{s} = 9$.
The solvent molecules undergo bounce-back collisions with the walls at $Z = 0$
and $L_Z$ that reverse their velocities.
Periodic boundary conditions are applied in the $X$ and $Y$ directions.

The catalytic (C) and noncatalytic (N) spheres in the dimer motors have
effective diameters $d_\text{C} = 4$ and $d_\text{N} = 8$, while the dimer mass
is $M_\text{m} = \frac{\pi}{6} \varrho_\text{s} m_\text{s} (d_\text{C}^3 +
d_\text{N}^3) \approx 2714$, which makes it neutrally buoyant.
In order to better visualize the collective dimer dynamics, their motion in the
$Z$ direction was suppressed by using two long-range 9--3~Lennard-Jones wall
potentials, $V_{\text{w}S}(\zeta) = (3\sqrt{3}/2) \epsilon_{\text{w}S}
\bigl[(\sigma_{\text{w}S}/\zeta)^9 - (\sigma_{\text{w}S}/\zeta)^3\bigr]$,
where $\zeta$ is the distance of a sphere to the wall, and the interaction
parameters are $\epsilon_{\text{w}S} = 5$ and $\sigma_{\text{w}S} = L_Z/2$.
These potentials largely confine the dimer dynamics to the $XY$ midplane of the
slab, although the fluid flow and concentration fields are three dimensional.
For a given number of motors $N_\text{m}$ and area fraction
$\phi=N_\text{m} A_\text{m}/A$, where $A_\text{m} = \frac{\pi}{4}
(d_\text{C}^2 + d_\text{N}^2)$ is the area of the dimer projected on the
$XY$ plane, and $A=L^2$, the edge lengths of the midplane are chosen as
$L = \big\lfloor\sqrt{A_\text{m} N_\text{m}/\phi}\big\rfloor$.
(The projected area could be modeled as that of a cylindrical representation of
the dimer and this would yield somewhat higher area fractions.)

Dimer spheres and solvent molecules, as well as spheres of different dimers,
interact via shifted, truncated 12--6~Lennard-Jones potentials of the form
$V_{ij}(r) = \epsilon_{ij}\bigl\{4\bigl[(\sigma_{ij}/r)^{12}
- (\sigma_{ij}/r)^6\bigr] + 1\bigr\}$
for $r < \sqrt[6]{2}\,\sigma_{ij}$ and zero otherwise.
The separation distances are $\sigma_\text{CA} = \sigma_\text{CB} =
d_\text{C}/2$ and $\sigma_\text{NA} = \sigma_\text{NB} = d_\text{N}/2$ for
sphere--solvent pairs, and $\sigma_\text{CC} = d_\text{C} + 1$,
$\sigma_\text{CN} = (d_\text{C} + d_\text{N})/2 + 1$, and $\sigma_\text{NN}
= d_\text{N} + 1$ for sphere--sphere pairs.
The interaction energies are $\epsilon_\text{CA} = \epsilon_\text{CB} =
\epsilon_\text{NA} = 1$ and $\epsilon_\text{NB} = 0.1$ or 10 for
sphere--solvent pairs and $\epsilon_\text{CC} = \epsilon_\text{CN} =
\epsilon_\text{NN} = 10$ for sphere--sphere pairs.
The dimer bond length is $d_\text{CN} = \sqrt[6]{2}\,(d_\text{C} + d_\text{N})/2$,
which yields the maximum possible propulsion velocity~\cite{reigh2015} at the
minimum possible distance that still ensures conservation of energy in the
presence of chemical reactions at the catalytic sphere.
Likewise, the sphere--sphere pair separation distances and interaction energies
given above are large enough to ensure conservation of energy in the presence
of chemical reactions and avoid the occurrence of solvent depletion forces
between dimers.%
~\footnote{The simulation code explicitly asserts at every MD step that any
solvent molecule interacts with at most one dimer sphere, which, besides
satisfying the sphere--sphere separation criteria, allows for a performance
optimization in the parallel computation of sphere--solvent forces.}
There are no intermolecular potentials among solvent molecules; these
interactions are instead described by multiparticle collision dynamics.

In hybrid MD--MPCD, the solvent particles undergo multiparticle collisions at
discrete time intervals $\tau_\text{MPC}$.
As described in detail
elsewhere~\cite{malevanets1999,*malevanets2000,kapral2008,gompper2009},
to carry out multiparticle collisions, at discrete times the system is
partitioned into collision cells, and rotation operators are assigned to the
cells.
Post-collision particle velocities are then obtained in each cell by rotating
the particle velocity relative to the cell center of mass velocity and adding
back the center of mass velocity.%
~\cite{[{MPCD collisions in a cell were carried out using rotations by $\pi/2$
about a randomly chosen axis.
Galilean invariance was guaranteed by implementing grid shifting when carrying
out the collisions; }]ihle2001,*ihle2003}
Between these multiparticle collisions, all particles in the system evolve by
Newton's equations of motion in the potentials given above.
Since MPCD conserves mass, momentum, and energy, one may derive the
Navier-Stokes equations for the solvent on long distance and time scales with
known values of the transport coefficients.%
~\cite{malevanets1999,*malevanets2000,ihle2001,*ihle2003,kapral2008,gompper2009}
The MPCD collision cell size is $a = 1$, and the solvent number density is
$n = 9$.
The velocity-Verlet time step is $\tau_\text{MD} = 0.01$, and the thermal
energy of the system is $k_\text{B}T = 1/6$.
Given these parameters, two values of the multiparticle collision time
$\tau_\text{MPC}$ are considered.
For $\tau_\text{MPC} = 0.5$, the solvent viscosity is $\eta = 1.22$, the A
and B common diffusion coefficient is $D_\text{A} = 0.099$ and the Schmidt
number is $\text{Sc} = 1.4$.
For $\tau_\text{MPC} = 0.1$, the solvent viscosity is $\eta = 4.51$, the A
and B common diffusion coefficient is $D_\text{A} = 0.020$ and the Schmidt
number is $\text{Sc} = 25$.
We note that for our collision model, the solvent molecules interact with the
motors through intermolecular potentials so that the fluid particle diffusion
does not enter in the Navier-Stokes equation.
In this circumstance, the fluid will still exhibit liquid-like properties in
spite of the relatively low values of the Schmidt numbers, which are a
consequence of coarse-graining to achieve computational efficiency.~%
\cite{padding2006}

The catalytic chemical reactions that drive the directed motion are carried out
by converting fuel to product in irreversible%
~\cite{[{While irreversible reactions are considered here, the reactive
dynamics may be generalized to treat reversible reactions.
See, for example, }]thakur2011}
reactions $\text{A} + \text{C}\rightarrow\text{B} + \text{C}$ with unit
probability in the vicinity of a catalytic sphere, $r_\text{CA} <
\sqrt[6]{2}\,\sigma_\text{CA}$.
Given this reactive dynamics, the rate constant for reactions on a single
catalytic sphere, $k$, may be approximately determined~\cite{tucci2004} and is
given by $k^{-1} = k_0^{-1}+k_D^{-1}$, where $k_0 = p_0 r_\text{C}^2 (8\pi
k_\text{B}T/m)^{1/2}$ is the intrinsic reaction rate coefficient and
$k_D = 4\pi D_\text{A} r_\text{C}$ is the Smoluchowski diffusion-controlled
rate coefficient, where the reaction probability is $p_0 = 1$, and $r_\text{C}$
is the radius of the catalytic sphere.
For our parameter values, $k_0 = 8.19$, and for $\tau_\text{MPC} = 0.5$, we
have $k_D = 2.48$ corresponding to diffusion-influenced kinetics, while for
$\tau_\text{MPC} = 0.1$, we have $k_D = 0.50$ so that the reaction is
diffusion-controlled.

In order to maintain the system in a steady state, irreversible reactions
$\text{B}\stackrel{k_2}{\rightarrow} \text{A}$ occur locally in the bulk of the
solution with rate coefficient $k_2 = 0.01$ outside the sphere--solvent
interaction zones $r_{S\text{B}} > \sqrt[6]{2}\,\sigma_{S\text{B}}$ for all
spheres $S$.
For this purpose, the reactive version of MPCD was employed.~\cite{rholf2008}
On long distance and time scales, this reactive dynamics yields the
reaction-diffusion equations.

All parameter values listed above, as well as the results that follow, are
reported in dimensionless units with mass in units of $m$, energy in units of
$\epsilon$, distances in units of $\sigma$, and time in units of
$t_0=\sqrt{m \sigma^2/\epsilon}$.

\section{Single sphere-dimer motor}\label{sec:single}
\begin{figure}[t]
  \setlength{\fboxsep}{0pt}%
  \subfloat[\label{fig:forward}$\epsilon_\text{NB} = 0.1$]{\fbox{\includegraphics[width=0.48\linewidth]{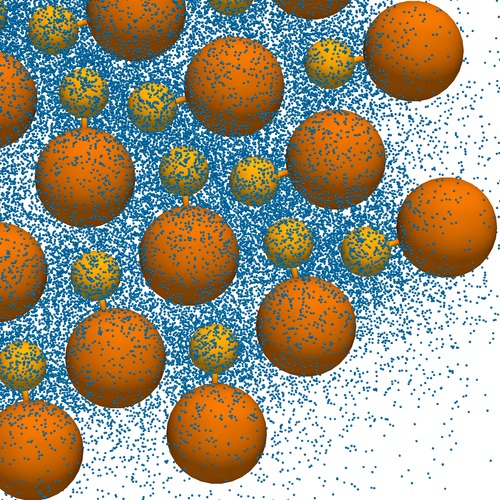}}}\hfill%
  \subfloat[\label{fig:backward}$\epsilon_\text{NB} = 10$]{\fbox{\includegraphics[width=0.48\linewidth]{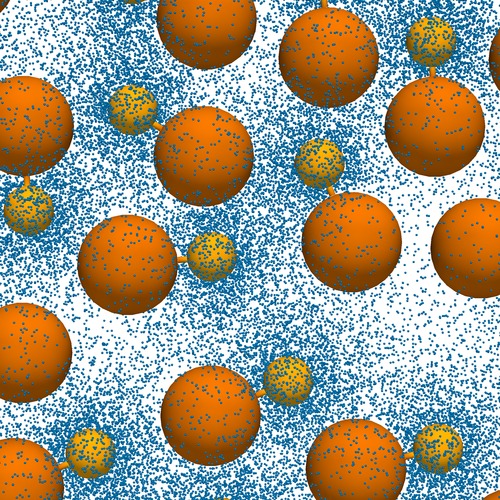}}}%
  \caption{Partial snapshots of (a) forward-moving and (b) backward-moving
  sphere-dimer motors comprising catalytic
  (\textcolor[rgb]{0.9,0.6,0.0}{orange}) and noncatalytic spheres
  (\textcolor[rgb]{0.8,0.4,0.0}{vermillion}) in a solvent consisting of fuel
  (not shown for clarity) and product molecules
  (\textcolor[rgb]{0.0,0.45,0.7}{blue}).}
\end{figure}

It is useful to summarize the properties of a single motor in solution before
studying the collective dynamics of many motors.%
~\footnote{The simulations of single motors were carried out using systems
of size $L_X = L_Y = 60$ and $L_Z = 30$, and averaging was performed over
60 realizations of $10^8$ MD steps each.}
The majority of the results presented below are for two values of
$\epsilon_\text{NB}$.
For $\epsilon_\text{NB} = 0.1$, the sphere dimer moves in a direction with the
catalytic sphere at its head, which for convenience we shall refer to as a
forward-moving motor (Fig.~\ref{fig:forward}), while for $\epsilon_\text{NB} =
10$, the motor has the noncatalytic sphere at its head and will be called a
backward-moving motor (Fig.~\ref{fig:backward}).
Letting $\hat{\bf z}$ be a unit vector directed along the dimer bond from the
N to C spheres, the projection of the motor velocity along $\hat{\bf z}$ will
be denoted by $V_z$.
The motor reorientation time, $\tau_\text{R}$, is determined from the decay of
the orientation correlation function
$\langle \hat{\bf z}(t)\cdot \hat{\bf z}\rangle$.
The time it takes a motor to move a distance equal to its effective radius,
$R_\text{m} = \sqrt[3]{d_\text{C}^3 + d_\text{N}^3}/2 = 4.16$, is the
ballistic time, $\tau_\text{B} = R_\text{m}/|V_z|$, while the time it takes
the motor to diffuse this distance is the diffusion time, $\tau_\text{D} =
R_\text{m}^2 / D_0$, where $D_0$ is the diffusion coefficient of an inactive
dimer.
P{\'e}clet numbers can be defined as the ratio of the diffusion and ballistic
times, $\text{Pe}^\prime = \tau_\text{D}/\tau_\text{B}$, and the ratio of the
orientational and ballistic times, $\text{Pe} = \tau_\text{R}/\tau_\text{B}$.
The values of these quantities can be found in Table~\ref{tab:single}.
\begin{table}[htbp]
  \centering
  \begin{ruledtabular}
    \begin{tabular}{lllllllll}
      & $\tau_\text{MPC}$ & $V_z$ & $D_0$ & $\tau_\text{D}$ & $\tau_\text{R}$
      & $\tau_\text{B}$ & $\text{Pe}$ & $\text{Pe}^\prime$ \\
      & & $\times 10^3$ & $\times 10^3$ & & & & & \\
      \hline
      f-m & 0.5 & 26   & 2    & 8241  & 3305  & 160 & 21  & 52 \\
      f-m & 0.1 & 8.1  & 0.78 & 22187 & 13677 & 514 & 27  & 43 \\
      \hline
      b-m & 0.5 & -11  & 2    & 8241  & 7362  & 378  & 19 & 22 \\
      b-m & 0.1 & -2.9 & 0.78 & 22187 & 16790 & 1434 & 12 & 15 \\
    \end{tabular}
  \end{ruledtabular}
  \caption{Properties of forward-moving (f-m) and backward-moving (b-m) single
    sphere-dimer motors for two different solvent conditions characterized by
    the values of $\tau_\text{MPC}$.}
  \label{tab:single}
\end{table}

The continuum theory of the sphere-dimer motor is most conveniently
formulated in a bispherical coordinate system.%
~\cite{stimson1926,morse1953,happel1973,reigh2015,michelin2015,popescu2011}
The concentration fields can be found by applying a radiation boundary
condition involving $k_0$ on the catalytic sphere and a reflecting boundary
condition on the noncatalytic sphere.
The motor velocity has a functional form typical for phoretic propulsion,
\begin{equation}\label{eq:Vcon_theory}
  V_z = \frac{k_\text{B} T}{\eta} \Lambda Q,
\end{equation}
where
\begin{equation}\label{eq:Lambda}
  \Lambda = \int_0^\infty r[e^{-\beta V_\text{NB}(r)}-e^{-\beta V_\text{NA}(r)}]dr.
\end{equation}
Here $\beta=1/(k_\text{B}T)$, $V_\text{NB}$ and $V_\text{NA}$ denote the
interaction potentials of the B and A species with the noncatalytic sphere,
respectively, while the factor $Q$ depends on the system parameters and is
determined from the solutions of the reaction-diffusion and Stokes equations in
bispherical coordinates.
The explicit form for the motor velocity is given in Eq.~(8) of
Ref.~\onlinecite{reigh2015} from which the factor $Q$ can be found.
The continuum solutions have been compared with microscopic simulations.
In particular, experiment~\cite{valadares2010}, continuum theory, and
simulation~\cite{reigh2015,tao2008} indicate that the maximum motor velocity
occurs near a 1:2 ratio of catalytic to noncatalytic sphere diameters, and this
has motivated the radius ratio chosen for most of the work in this study,
although results for sphere-dimer motors made from spheres with equal sizes
will also be given.
The continuum results for the fluid velocity fields may also be obtained
analytically and depend on the dimer bond length.
For the bond length chosen in this study, the forward-motor far-field flow
decays as $1/r^2$, characteristic of a force dipole, and the flow pattern is
that of a ``puller'' where the fluid flow is inward from the front and rear of
the motor and outward on its sides (cf. Fig.~4 of Ref.~\onlinecite{reigh2015}).
The near-field flow pattern is more complex with fluid circulation that changes
the puller character.
The flow pattern is reversed for the backward motor with a far-field flow
characteristic of a ``pusher''.

\section{Collective dynamics}\label{sec:collective}
In this section, we describe the dynamics at selected points in the system
parameter space.
The nature of the collective behavior depends on the values of the microscopic
parameters that define the system: intermolecular potentials, MPCD parameters
for the solvent, and motor density.
Once these parameters and system conditions are specified, all properties
of the system may be determined from the solution of the evolution equations.%
~\cite{[{The simulations were performed using a code written in OpenCL~C and
Lua that is published under a free software license; }]colberg2013}
Unless stated otherwise, results will be reported for $\tau_\text{MPC} = 0.5$.
It is useful to organize the discussion of the collective dynamics based on
whether self-propulsion leads to forward or backward motor motion.

\subsection*{Forward-moving motors}
Consider interaction energies $(\epsilon_\text{NB} = 0.1,\epsilon_\text{NA}
= 1)$ where the parameter $\Lambda = 1.18 > 0$ (see Eq.~\eqref{eq:Lambda}) and
the motor moves in the forward direction with the C sphere at its head.
Since the catalytic reaction produces a high concentration of species B and
depletes that of A near the portion of the N sphere surface closest to the C
sphere, for these potential parameters, a net force will act to move the motor
in the forward direction.
It also follows from this that a motor will respond to the concentration
gradients of other motors by moving in the direction of high B concentration,
i.e., forward-moving motors are chemotactically attracted to each other.
An individual motor will respond to both its self-generated concentration
gradient and that due to other motors, and forward-moving motors tend
to align with their catalytic heads pointing in the same direction.
Geometric factors related to the motor shape also play a role in such
alignment processes since they influence how the motors interact and assemble
when they are close to each other.

Hydrodynamic interactions also contribute to collective dynamics and these
effects have been investigated for active swimmers.
The far-field flow of pullers is attractive in the swimming direction and
repulsive in the perpendicular directions, giving rise to a tendency for two
swimmers to move away from each other.~\cite{lauga2009}
Simulations of suspensions of pullers do not show large-scale correlated
motions~\cite{saintillan2012}, but when near-field flows and the finite volume
of swimmers are taken into account, large density fluctuations have been observed
for pullers.~\cite{oyama2016,alarcon2013}
Recall that on the basis of the far-field flow, for our dimer bond length,
the forward-moving motor can be classified as a puller, but the near-field flow
is more complex.

\begin{figure}[htbp]
  \centering
  \includegraphics{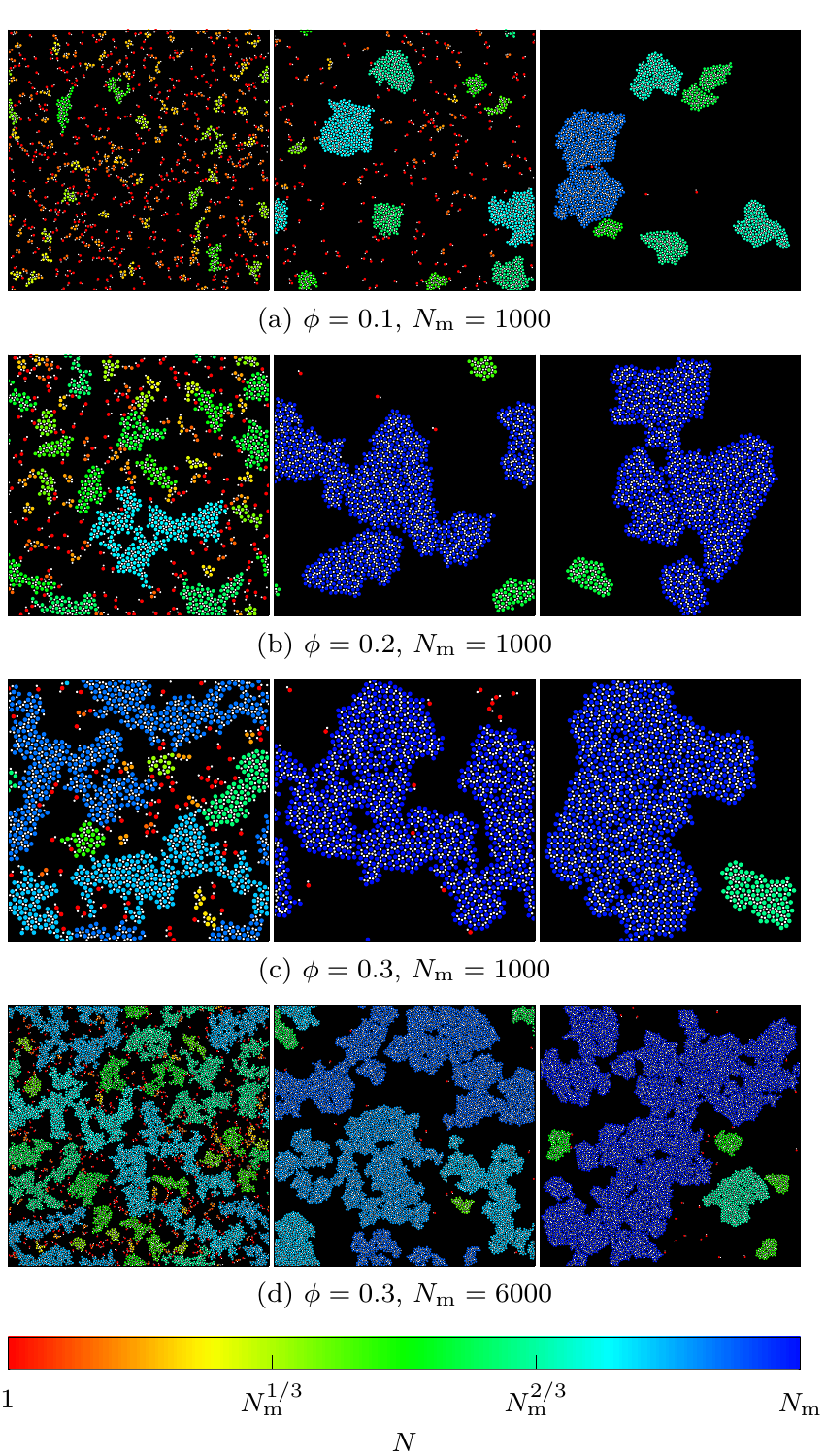}
  \caption{Instantaneous configurations of $N_\text{m} = 1000$ and 6000
    forward-moving motors for different area fractions $\phi$ at short,
    intermediate, and long times (from left to right: $t = \num{e4}$,
    \num{e5}, and \num{5e5}).
    Clusters of $N$ motors are colored according to a logarithmic scale
    shown below the figure, where a pair of motors is considered the same
    cluster if $V_{SS^\prime}(r) \neq 0$ for spheres $S$ and $S^\prime$.}
  \label{fig:forward_cluster}
\end{figure}
The evolution of systems with $N_\text{m} = 1000$ and 6000 sphere-dimer
motors for several area fractions $\phi$ is shown in Fig.~\ref{fig:forward_cluster}.
The number of solvent particles $N_s$ in the system ranges from $N_s\approx
10^7$ to $10^8$ depending on the area fraction.
The initial state of the system was prepared by randomly placing the dimers
with random orientation and without overlap such that $V_{SS^\prime}(r) = 0$
for spheres $S$ and $S^\prime$ and with the solvent comprising only A
molecules.
Segregation into low-density gas-like and high-density disordered solid-like
phases was observed, with the time scale for coarsening dependent on the area
fraction.
A video showing the evolution of clusters can be found in the supplementary
material.
For the area fractions considered, dimer motors propagate with a strong
ballistic component between encounters.
Cluster growth occurs through single dimer attachment to a cluster and by
cluster--cluster aggregation events involving propagating clusters.
Due to the presence of thermal noise, clusters evaporate, fragment, and
combine, although in the late stages where large clusters exist, the gas phase
is very dilute and the time scales for cluster breakup and formation are very
long.

\begin{figure}[htbp]
  \centering
  \subfloat[\label{fig:forward_cluster_size}]{\includegraphics{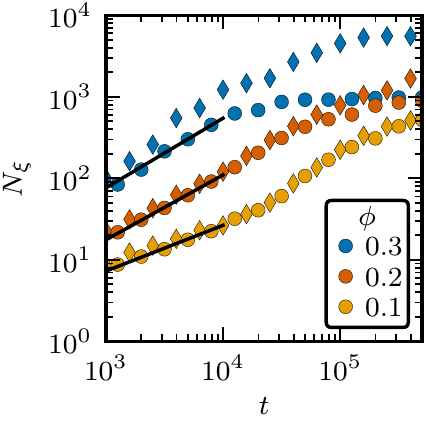}}\hfill%
  \subfloat[\label{fig:forward_cluster_exponent}]{\includegraphics{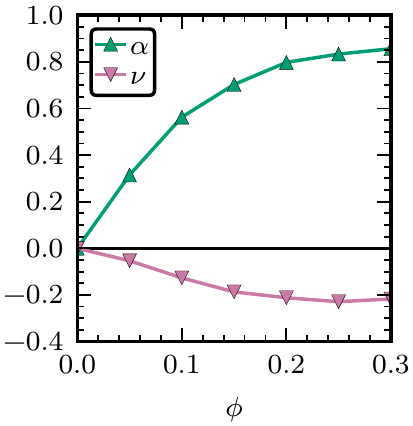}}%
  \caption{(a) Average number of forward-moving motors in the largest cluster,
    $N_{\xi}$, versus time for three area fractions $\phi$.
    The circular markers show averages over 20 realizations of systems with
    $N_\text{m} = 1000$ motors.
    The diamond-shaped markers show averages over 3 realizations (to compensate
    for the increased computational cost) of systems with $N_\text{m} =
    2000$, 4000, and 6000 motors for $\phi = 0.1$, 0.2, and 0.3, respectively.
    The solid black lines show fits to power-laws in the initial time regimes.
    (b) Power-law exponents versus area fraction $\phi$ obtained from fits to
    the simulation data in the initial time regime: $\alpha$ for the size of
    the largest cluster (top curve) and $\nu$ for the number of motors in the
    gas phase (bottom curve).}
\end{figure}
The average number of dimers in the largest cluster, $N_\xi$, is plotted
as a function of time for different values of $\phi$ in
Fig.~\ref{fig:forward_cluster_size}.
For each area fraction, we compare the results for two different values of the
total number of motors in the simulation volume to gauge the magnitude of
finite-size effects.
The log--log scale plots of $N_\xi$ indicate power-law scaling of the cluster
growth that is dependent on $\phi$.
In the initial stage of the growth, for $t\leq 10^4$, $N_\xi(t) \propto
t^\alpha$.
For $\phi = 0.1$ and 0.2, the initial regime does not exhibit finite-size
effects; however, for $\phi = 0.3$, one can see that the initial regime for
$N_\text{m} = 1000$ is suppressed compared to that for $N_\text{m} = 6000$.
The cluster structure in this regime can be seen in the panels in the left-most
column in Fig.~\ref{fig:forward_cluster}.
The exponent $\alpha(\phi)$ was fitted for systems with $N_\text{m} = 1000$
and increases with $\phi$ (see Fig.~\ref{fig:forward_cluster_exponent}).
As above, finite-size effects may alter the exponent values for $\phi\ge 0.2$.
The simulations results for 4000 and 6000 dimers suggest the presence of
another power-law regime with smaller exponents at longer times, but more
extensive simulations of even larger systems are needed to verify its presence
and quantitatively characterize its behavior.

The temporal variation of the average number of motors in the gas phase,
$N_\text{g}$, also displays power-law behavior, $N_\text{g}(t)
\propto t^\nu$, in the initial time regime.
The value of the exponent $\nu(\phi)$ decreases with increasing $\phi$ as shown
in Fig.~\ref{fig:forward_cluster_exponent}.
This slow decay in the short-time regime is followed by a rapid decay to very
small values since the gas phase is very dilute.

Since solid-like clusters are quickly formed, the longtime dynamics of
individual motors is determined by diffusive motions of the clusters containing
these motors.
This diffusive cluster dynamics occurs on very long time scales and requires
significantly longer simulation times than those considered in this study in
order to determine the diffusion coefficients accurately.

\begin{figure}[htbp]
  \centering
  \includegraphics{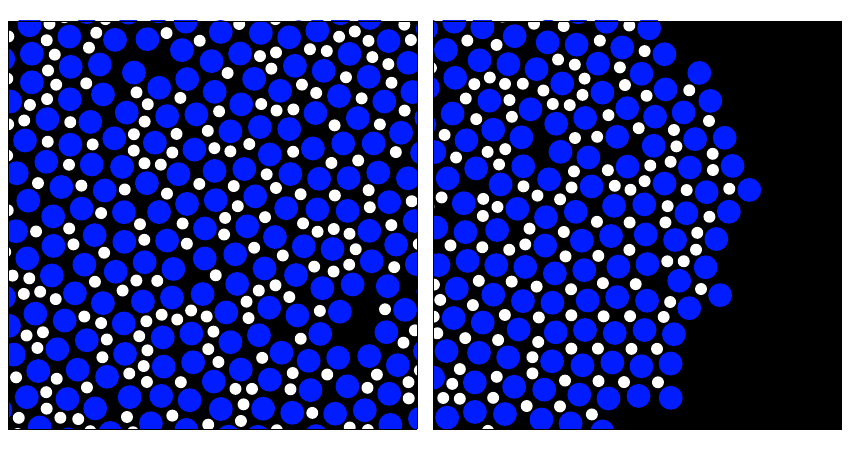}
  \caption{Instantaneous configuration of $N_\text{m} = 1000$ forward-moving
    motors for $\phi = 0.3$ at long times ($t = \num{5e5}$).
    The two panels show the interior of a cluster with a disordered lattice
    structure (left) and its boundary with a square lattice structure (right).}
  \label{fig:forward_cluster_lattice}
\end{figure}
In the steady state, large metastable clusters persist for very long periods of
time.
The structure of these disordered solid-like clusters is determined by the
interactions among motors, dependent on the sphere-dimer geometry, and the
chemotactic interactions among motors.
Fig.~\ref{fig:forward_cluster_lattice} shows expanded views of the interior and
periphery of a cluster.
There is a high degree of orientational order on the periphery with motors
pointing with their catalytic heads towards the bulk of the cluster.
This is a consequence of the attraction of these forward-moving motors to
regions of high product (B) concentration, which is concentrated in the cluster
vicinity, along with geometric effects that arise from the packing of the
asymmetric dimers into clusters.
The interior is globally disordered but has a high degree of local structural
order.
In the peripheral region, and elsewhere in the interior of the cluster, there
exist square lattice arrangements of N spheres with C spheres occupying the
interstitial positions.
Since the C and N spheres are linked, there is a high degree of dimer
orientational order.
Other prominent types of local ordering can be seen in the expanded views.
In particular, there are configurations in the cluster interior where dimers
are oriented so that their C heads are aligned in order to partially surround
an N~sphere.

\begin{figure}[htbp]
  \centering
  \subfloat[\label{fig:forward_rdf}]{\includegraphics{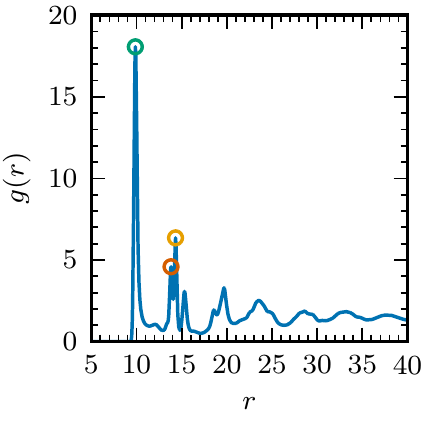}}%
  \subfloat[\label{fig:forward_orientation}]{\includegraphics{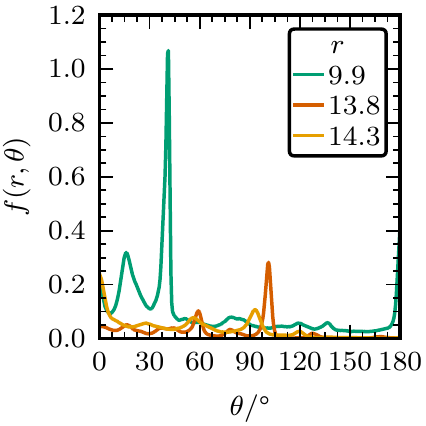}}%
  \caption{(a) Radial distribution function $g(r)$ of N--N sphere pairs
    for forward-moving motors at $\phi = 0.3$.
    (The scaled $g^\prime(r) = \phi\,g(r)$ for different $\phi$, not shown
    here, exhibit identical structure.)
    (b) Joint probability distribution $f(r, \theta)$ versus angle $\theta$
    between the bond vectors of pairs of motors at prominent peaks of $g(r)$.}
\end{figure}

These qualitative observations are reflected in the forms of the radial and
orientational correlations.
The NN radial distribution function is defined as
\begin{equation}
  g(r) = \frac{1}{2\pi r n_\text{N}}\Big\langle
  \sum_{j<i=1}^{N_\text{N}}\delta(|{\bf r}_{\text{N}ij}|-r)\Big\rangle \;,
\end{equation}
with ${\bf r}_{\text{N}ij} = {\bf r}_{\text{N}i}-{\bf r}_{\text{N}j}$,
where ${\bf r}_{\text{N}i}$ is the position of the noncatalytic monomer $i$,
$N_\text{N}$ is the number of noncatalytic monomers, and
$n_\text{N} = N_\text{N}/V$.
The angle brackets denote an average over time and realizations.
This function is plotted in Fig.~\ref{fig:forward_rdf}.
There is a very prominent peak at $r=9.9$ that corresponds to the NN packing
discussed above, with smaller-amplitude weakly damped oscillations that persist
for long distances reflecting the approximate long-range order.
Fine structure is seen in some of the smaller-amplitude next-neighbor peaks.
Fig.~\ref{fig:forward_orientation} shows the joint probability distribution
$f(r, \theta)$ of N--N sphere pairs,
\begin{equation}
  f(r,\theta)=\frac{1}{N_\text{N}}\Big\langle
  \sum_{j<i=1}^{N_\text{N}}\delta(\theta_{ij}-\theta)
  \delta(|{\bf r}_{\text{N}ij}|-r)\Big\rangle \;,
\end{equation}
where $\hat{{\bf z}_i}\cdot \hat{{\bf z}_j}=\cos \theta_{ij}$ and $\theta_{ij}$
is the angle between the unit bond vectors centered on noncatalytic spheres $i$
and $j$ for values of $r$ corresponding to some of the prominent peaks in
$g(r)$, namely, those at $r = 9.9$, 13.8, and 14.3.
In particular, note that strong orientational order is associated with the peak
at $f(r=9.9,\theta=40)$.
This peak arises from configurations in the cluster where nearest-neighbor
dimers are oriented with their heads pointing towards a central noncatalytic
sphere.

\subsection*{Backward-moving motors}
For interaction energies $(\epsilon_\text{NB} = 10, \epsilon_\text{NA} = 1)$,
where $\Lambda = -0.60 < 0$, the motor will move in the backward direction with
the N~sphere at its head.
Backward motors tend to move in a direction of lower product concentration and
will chemotactically respond to other motors by moving away from them, although
geometric factors play a role and aligned pairs of motors exist and move in the
backward direction as a unit for short periods of time.
As a result, one might expect that clustering will be less pronounced for such
motors.
This is what is found, but transient clustering has important effects on the
collective dynamics of these motors.

\begin{figure}[htbp]
  \centering
  \includegraphics{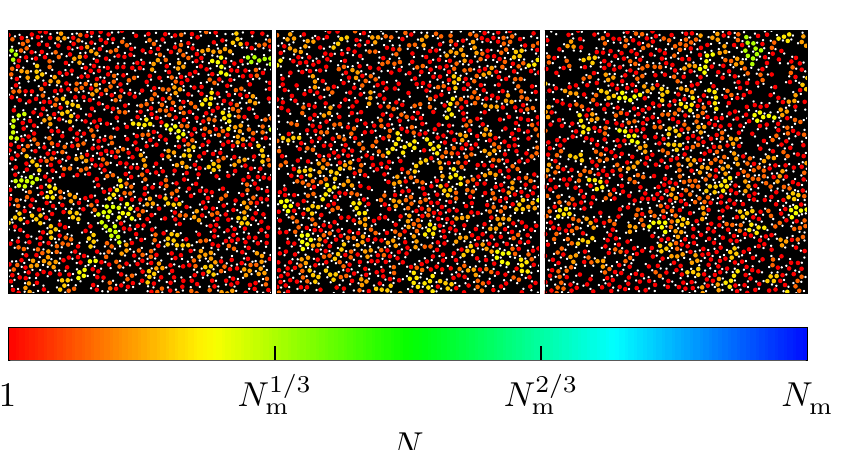}
  \caption{Instantaneous configurations of $N_\text{m} = 1000$ backward-moving
    motors for $\phi = 0.3$ at short, intermediate, and long times (from left
    to right: $t = \num{e4}$, \num{e5}, and \num{5e5}), showing the existence
    of a statistically stationary regime.
    Clusters of $N$ motors are colored to a logarithmic scale.}
  \label{fig:backward_cluster}
\end{figure}
\begin{figure}[htbp]
  \centering
  \subfloat[\label{fig:backward_rdf}]{\includegraphics{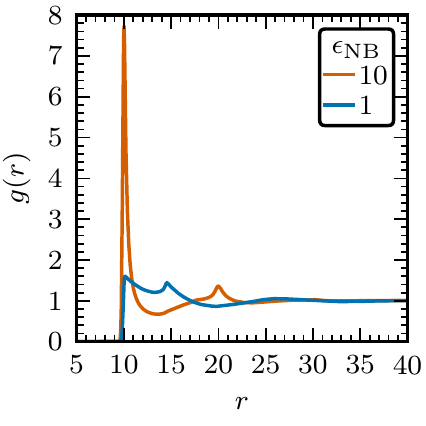}}%
  \subfloat[\label{fig:backward_compressibility}]{\includegraphics{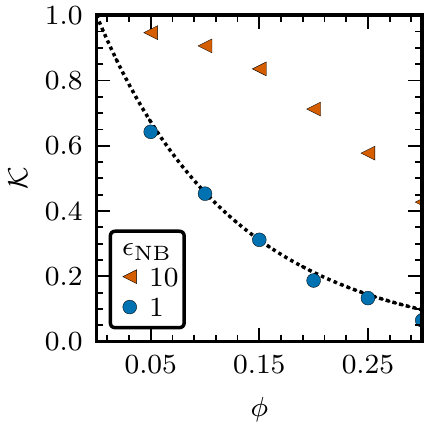}}%
  \caption{(a) Radial distribution function $g(r)$ of N--N sphere pairs
    for backward-moving motors and inactive dimers at area fraction
    $\phi = 0.3$.
    ($g(r)$ for different $\phi$, not shown here, are qualitatively similar
    with less pronounced peaks for decreasing $\phi$.)
    (b) ${\mathcal K}$ for backward-moving motors and inactive dimers versus
    area fraction $\phi$.}
\end{figure}

Fig.~\ref{fig:backward_cluster} shows instantaneous configurations of
backward-moving motors at several times.
No large-scale cluster formation takes place as for forward-moving motors;
however, substantial regions of inhomogeneous density are evident in the
figure.
(See also the video in the supplementary material.)
The tendency to form small transient clusters can be quantified by considering
the steady state radial distribution functions, $g(r)$, with $r$ being the
magnitude of the distance between the N~spheres.
The distribution functions for inactive and backward-moving motors are
compared in Fig.~\ref{fig:backward_rdf}.
There is a strong peak in the NN distribution function at $r = 10$ for
backward-moving motors, while only very weak structural ordering is seen for
inactive dimers.

The tendency to form small transient clusters is reflected in the $\phi$
dependence of ${\mathcal K}$, defined as
\begin{equation}
 {\mathcal K} =
  1 + 2\pi\rho_\text{m} \int_0^{\infty}dr\,r\left(g(r)-1\right)\,,
\end{equation}
which is the nonequilibrium analog of ${\mathcal K}_\text{T}=
k_\text{B}T\rho_\text{m}\,\kappa_\text{T}$, where $\kappa_\text{T}$ is the
isothermal compressibility.
However, ${\mathcal K}$ depends on the nonequilibrium steady state radial
distribution function instead of its equilibrium analog.
Fig.~\ref{fig:backward_compressibility} plots ${\mathcal K}$ as a function of
$\phi$ for active and inactive dimers.
For inactive dimers, ${\mathcal K}$ decays quickly with increasing area fraction.
Its behavior is similar to that of a fluid of rigid hard spheres.
Using the Carnahan-Starling equation of state~\cite{carnahan1969} for a hard
sphere fluid, ${\mathcal K}_\text{HS}$ is given by
\begin{equation}
  {\mathcal K}_\text{HS} =
  \left(1 + \frac{8 \phi-2\phi^2}{(1-\phi)^4} \right)^{-1}.
\end{equation}
This quantity is also plotted in Fig.~\ref{fig:backward_compressibility} and
matches the inactive sphere dimer result.
In contrast to inactive dimers, ${\mathcal K}$ for backward-moving motors is
larger and has a convex rather than a concave shape.
For equilibrium systems, $\kappa_\text{T}$ is related to the number
fluctuations by $k_\text{B}T\rho_\text{m}\,\kappa_\text{T} =
(\langle N^2\rangle-\langle N\rangle^2)/\langle N\rangle$, and the markedly
different $\phi$ dependence of this function for active dimers reflects the
presence of strong density fluctuations.
For low area fractions, all of these expressions tend to the ideal gas value of
unity.

\begin{figure}[htbp]
  \centering
  \includegraphics{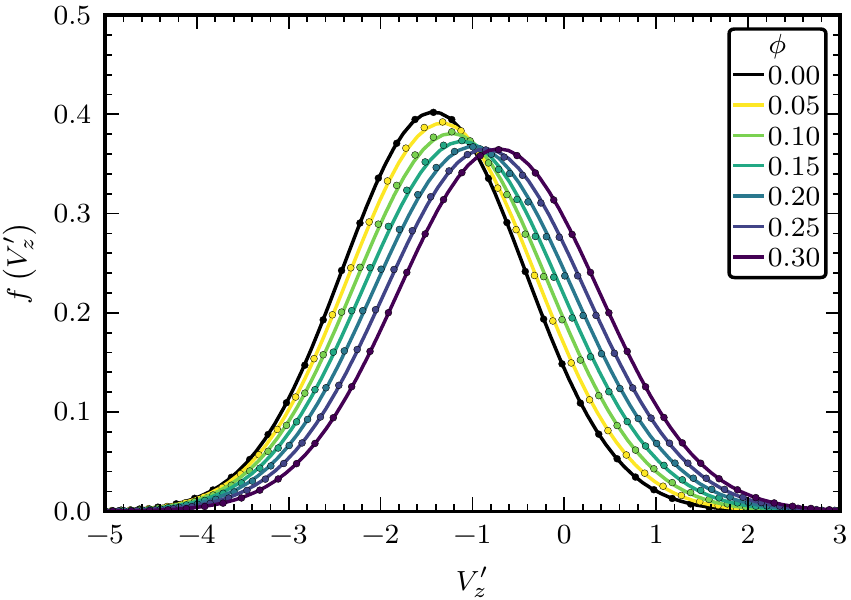}
  \caption{Steady-state velocity distribution $f(V_z^\prime)$ of scaled
    propulsion velocity $V_z^\prime = V_z / \sqrt{k_\text{B}T/M_\text{m}}$
    for backward-moving motors for different area fractions $\phi$.
    The points indicate the measured probability distributions.
    The lines show Gaussian distributions with the measured mean
    $\langle V_z\rangle$ and variance $\sigma_{V_z}^2$.}
  \label{fig:backward_velocity}
\end{figure}
Interactions among motors reduce the mean propulsion velocity of a motor as
seen in Fig.~\ref{fig:backward_velocity}, which plots the
probability distributions of the velocity projected along the propagation
direction for various values of $\phi$.
The distributions are Gaussian with mean values corresponding to the propulsion
velocity and widths close to those for a thermal distribution of velocities.
(The velocity is scaled by its thermal value,
$\sqrt{k_\text{B}T/M_\text{m}}$.)
The magnitude of the mean motor velocity steadily decreases and its width
increases as $\phi$ increases.
(See data in Table~\ref{tab:backward}.)
\begin{table}[htbp]
  \centering
  \begin{ruledtabular}
  \begin{tabular}{lllllll}
    $\phi$ & $\langle V_z^\prime\rangle$ & $\sigma_{V_z^\prime}$ & $D_\text{m}$
    & $D_0$ & $\tau_\text{R}$ & $\tau_\text{R}^0$ \\
    & & & & $\times 10^3$ & & \\
    \hline
    0    & -1.42 & 0.99 & 0.54 & 2.0 & 7362 & 5691 \\
    0.05 & -1.33 & 1.02 & 0.30 & 1.8 & 5542 & 5792 \\
    0.1  & -1.23 & 1.05 & 0.22 & 1.7 & 4759 & 6020 \\
    0.15 & -1.11 & 1.07 & 0.17 & 1.5 & 4393 & 6312 \\
    0.2  & -0.99 & 1.09 & 0.13 & 1.3 & 4264 & 6761 \\
    0.25 & -0.86 & 1.09 & 0.10 & 1.2 & 4313 & 7465 \\
    0.3  & -0.71 & 1.09 & 0.07 & 1.0 & 4558 & 8593 \\
  \end{tabular}
  \end{ruledtabular}
  \caption{Mean and standard deviation of scaled propulsion velocity
    $V_z^\prime = V_z / \sqrt{k_\text{B}T/M_\text{m}}$, effective
    diffusion coefficient $D_\text{m}$, and reorientation time
    $\tau_\text{R}$ for backward-moving motors, and diffusion coefficient $D_0$
    and reorientation time $\tau_\text{R}^0$ for inactive dimers, for different
    area fractions $\phi$.}
  \label{tab:backward}
\end{table}

\begin{figure}[htbp]
  \centering
  \subfloat[\label{fig:backward_reorientation}]{\includegraphics{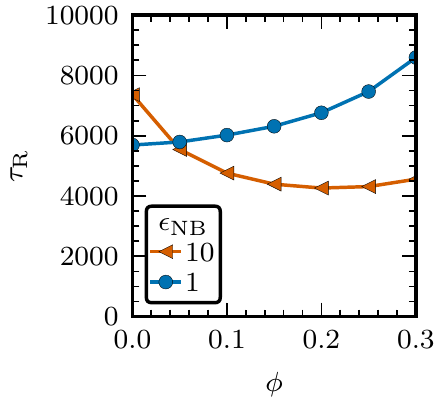}}\hfill%
  \subfloat[\label{fig:backward_diffusion}]{\includegraphics{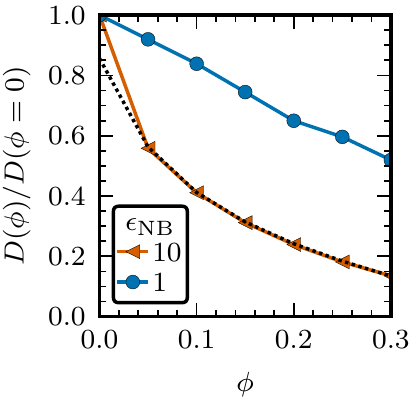}}%
  \caption{(a) Reorientation time $\tau_\text{R}$ for backward-moving motors
    and inactive dimers versus area fraction $\phi$.
    (b) Diffusion constants of backward-moving motors and inactive dimers
    versus area fraction $\phi$.
    The values are scaled to the respective single-motor diffusion constants
    ($\phi=0$).
    The dotted line shows the theoretical estimate of $D_\text{m}(\phi)$ from
    Eq.~\eqref{eq:Dm}.}
\end{figure}

The effects of crowding on dimer orientational dynamics are very different for
active motors and inactive dimers.
Fig.~\ref{fig:backward_reorientation} plots the reorientation time, $\tau_\text{R}$,
determined from the decay of the orientation autocorrelation function, for both
of these cases as function of the dimer area fraction.
For inactive dimers that execute thermal orientational Brownian motion, $\tau_\text{R}$
is an increasing function of $\phi$.
This is the expected effect due to crowding that will tend to hinder
reorientation.
Active dimers undergo ballistic motion in a direction along the dimer bond
vector.
Collisions with other motors at low dimer densities will perturb the ballistic
dynamics and shorten the reorientation time.
However, as the dimer density increases, dimers tend to form oriented transient
clusters, and this increases the reorientation time.
As a result, $\tau_\text{R}$ is a nonmonotonic function of $\phi$.
For small ensembles of forward-moving motors, $\tau_\text{R}$ was observed to
vary in a similar nonmonotonic fashion with dimer number.~\cite{thakur2012}
\begin{figure}[htbp]
  \centering
  \includegraphics{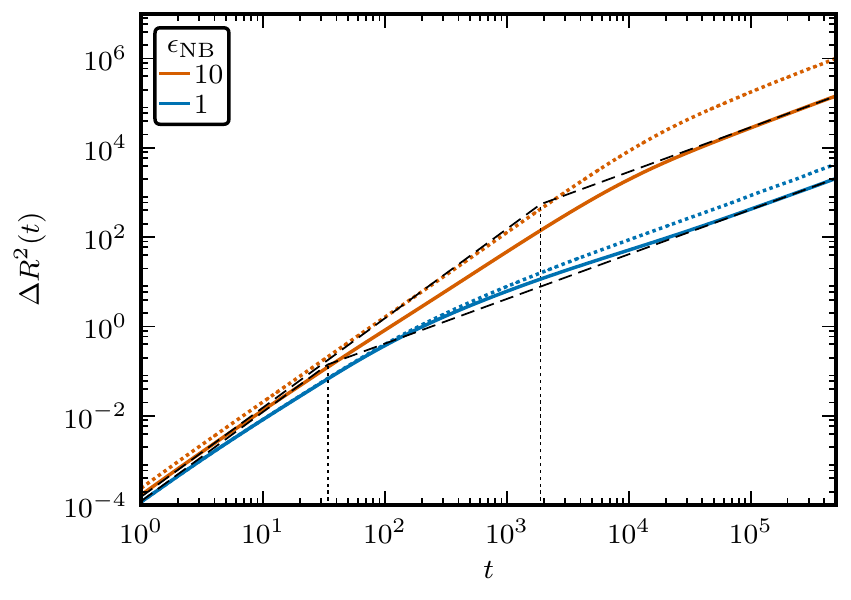}
  \caption{Mean-square displacement $\Delta R^2(t)$ of backward-moving motors
    and inactive dimers for $\phi = 0.3$.
    The dotted lines indicate the corresponding $\Delta R^2(t)$ for a single
    motor.
    The dashed lines show the ballistic and diffusive regimes at short and long
    times, respectively.
    The vertical lines mark the crossover time from the ballistic to the
    diffusive regime.}
  \label{fig:backward_msd}
\end{figure}

The existence of transient clustering is also reflected in the self-diffusion
coefficients of backward-moving motors.
For a single motor in solution, the mean square displacement $\Delta R^2(t)$
takes the form
\begin{eqnarray}\label{eq:MSD}
  \Delta R^2(t) & &= 4D_\text{m}t
  - 2V_z^2\tau_\text{R}^2\left(1-{\mathrm e}^{-t/\tau_\text{R}}\right)
  \nonumber \\
  && - 4\frac{k_\text{B}T}{M_\text{m}}
  \tau_\text{v}^2\left(1-{\mathrm e}^{-t/\tau_\text{v}}\right),
\end{eqnarray}
where $\tau_\text{v}$ is the velocity relaxation time.
In the ballistic regime, $t\ll\tau_\text{v}$, $\Delta R^2(t) \approx
(2k_\text{B}T/M_\text{m} + V_z^2)\,t^2$, while for long times, $\Delta R^2(t)
\approx 4 D_\text{m}t$ with an effective sphere-dimer diffusion coefficient
of $D_\text{m} = D_0 + \frac{1}{2}V_z^2\tau_\text{R}$, where the bare diffusion
coefficient is $D_0 = (k_\text{B}T/M_\text{m})\,\tau_\text{v}$.
For a single backward-moving motor, the effective diffusion coefficient has a
value $D_\text{m} = 0.299$, while the bare diffusion coefficient is $D_0 = 0.002$.

Fig.~\ref{fig:backward_msd} compares $\Delta R^2(t)$ versus time for inactive
dimers and backward-moving motors, with the ballistic and diffusive regimes
indicated by the straight line segments that intersect at the crossover time
separating these regimes.
For a single dimer where Eq.~\eqref{eq:MSD} applies, the crossover time is
approximately $t_c \approx 2\tau_\text{R}$.
The inertial regime for inactive dimers is given by $\Delta R^2(t) \approx
2k_\text{B}T/M_\text{m}\,t^2$, and at $\phi = 0.3$, the crossover time
is $t_c \approx 30$, which is much shorter than $2\tau_\text{R}(\phi)$.
In addition, sub-diffusive dynamics is observed before the diffusive regime is
reached.
For backward-moving motors at $\phi = 0.3$, the sub-diffusive regime is
absent and the crossover time is $t_c \approx 2000$, which is still
considerably smaller than $2\tau_\text{R}(\phi) \approx 8000$.

The effective motor diffusion coefficient as a function of the area fraction
$\phi$, $D_\text{m}(\phi)$, was determined from the long-time behavior of
$\Delta R^2(t)$, and the ratio $D_\text{m}(\phi) / D_\text{m}(\phi=0)$ is
plotted versus $\phi$ in Fig.~\ref{fig:backward_diffusion}.
One observes that this ratio is smaller and decreases more rapidly for
backward-moving motors than for inactive dimers.
Although the diffusion coefficients of the active backward-moving motors are
much larger than the corresponding diffusion coefficients of inactive dimers,
the effects of crowding are more significant since the motor velocity and
reorientation time, which determine the effective diffusion coefficients of the
motors, are strong functions of $\phi$ (Table~\ref{tab:backward}).

\subsection*{Other motor and system parameters}
The phenomena observed for asymmetric sphere-dimer motors with $(d_\text{N} = 8,
d_\text{C} = 4)$, for the specific solvent conditions and interaction
potentials used above, are present in systems with other motor and solvent
parameters.
While we have not carried out a systematic study as a function of all
parameters that determine the dynamics, a few examples given below will serve
to illustrate the results.

\begin{figure}[htbp]
  \centering
  \includegraphics{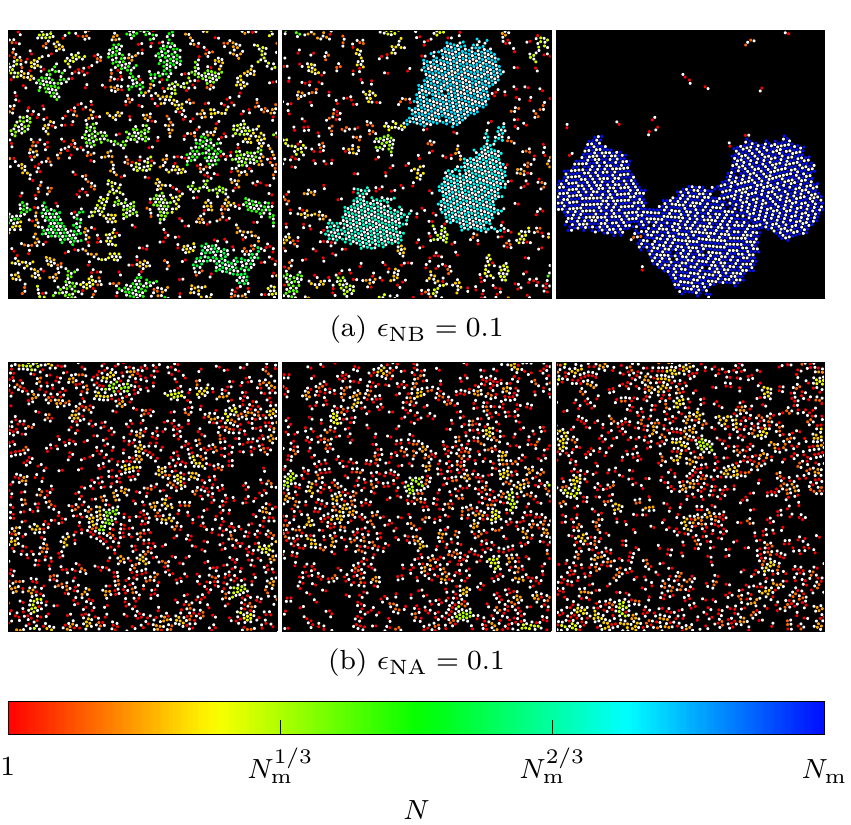}
  \caption{Instantaneous configurations of $N_\text{m} = 1000$ forward-moving
    (top) and backward-moving motors (bottom) with diameters $d_\text{C} =
    d_\text{N} = 8$ for $\phi = 0.2$ at short, intermediate, and long times
    (from left to right: $t = \num{e4}$, \num{e5}, and \num{5e5}).
    Clusters of $N$ motors are colored to a logarithmic scale.}
  \label{fig:symmetric_cluster}
\end{figure}

The interaction parameters $(\epsilon_\text{NB}, \epsilon_\text{NA})$
determining forward and backward motions were chosen to be $(0.1, 1)$ and
$(10, 1)$, respectively, corresponding to $\Lambda = 1.18$ and
$\Lambda = -0.60$ in the continuum theory expression for the motor velocity in
Eq.~(\ref{eq:Vcon_theory}).
We may also choose parameter pairs $(0.1, 1)$ and $(1, 0.1)$ so that
$\Lambda$ takes the values $\Lambda = \pm 1.18$.
The results for this more symmetric case again show clustering for forward
motors and only strong fluctuations for backward motors.
In addition, similar results, shown in Fig.~\ref{fig:symmetric_cluster}, are
found for symmetric sphere-dimer motors with $d_\text{N} = d_\text{C} = 8$
(see videos in the supplementary material).
The change to a symmetric motor geometry increases the tendency of
backward-moving motors to align, and transient strings of motors exist and play
a role in enhancing density fluctuations due to transient cluster formation.
For forward-moving motors, this tendency to form transient strings is less
significant since noncatalytic spheres are chemotactically attracted to
catalytic spheres.
The large clusters that form for forward-moving motors have hexagonal ordering
with defects.~\cite{thakur2012}
Additional information concerning these cases can be found in
Table~\ref{tab:modified}.

\begin{table*}[htbp]
  \centering
  \begin{ruledtabular}
    \begin{tabular}{lllllllllllllll}
      & $d_\text{C}$ & $d_\text{N}$ & $\tau_\text{MPC}$ & $\tau_\text{AT}$
      & $\epsilon_\text{NA}$ & $\epsilon_\text{NB}$ & $F_\text{m}$
      & $V_z$ & $D_0$ & $\tau_\text{D}$ & $\tau_\text{R}$ & $\tau_\text{B}$
      & $\text{Pe}$ & $\text{Pe}^\prime$ \\
      & & & & & & & $\times 10^3$ & $\times 10^3$ & & & & & \\
      \hline
      f-m & 4 & 8 & 0.5 & --  & 1   & 0.1 & -- & 26    & 2     & 8639    & 3305    & 159   & 20  & 54  \\
      b-m & 4 & 8 & 0.5 & --  & 0.1 & 1   & -- & -21   & 2     & 8639    & 8020    & 194   & 41  & 44  \\
      f-m & 8 & 8 & 0.5 & --  & 1   & 0.1 & -- & 25    & 1.5   & 16817   & 15517   & 203   & 76  & 82  \\
      b-m & 8 & 8 & 0.5 & --  & 0.1 & 1   & -- & -19   & 1.5   & 16817   & 17073   & 266   & 63  & 62  \\
      \hline
      f-m & 4 & 8 & --  & 0.5 & 1   & 0.1 & -- & 0.63  & 0.028 & 618576  & 165801  & 6649  & 24  & 93  \\
      b-m & 4 & 8 & --  & 0.5 & 0.1 & 1   & -- & -0.78 & 0.028 & 618576  & 135084  & 5357  & 25  & 115 \\
      f-m & 4 & 8 & --  & 0.5 & 1   & --  & 4  & 0.7   & 0.028 & 618576  & 167372  & 5933  & 28  & 104 \\
      b-m & 4 & 8 & --  & 0.5 & 1   & --  & -4 & -0.7  & 0.028 & 618576  & 152142  & 5961  & 25  & 103 \\
    \end{tabular}
  \end{ruledtabular}
  \caption{Properties of forward-moving (f-m) and backward-moving (b-m) single
    sphere-dimer motors for full microscopic model (top) and for two modified
    microscopic models (bottom) that neglect hydrodynamics using a global
    Andersen thermostat.}
  \label{tab:modified}
\end{table*}

Finally, for both forward- and backward-moving motors, the qualitative features
of the collective dynamics for the higher viscosity fluid with $\tau_\text{MPC}
= 0.1$ are similar to those described above for $\tau_\text{MPC} = 0.5$.
In particular, the forward-moving motors form large clusters, although the time
scale on which this cluster formation occurs is longer for $\tau_\text{MPC} = 0.1$
due to the smaller single-motor velocity (see Table~\ref{tab:single}).
Similarly, for the backward-moving motors, only transient cluster formation is
observed.

\section{Factors contributing to collective behavior}\label{sec:factors}
Segregation into domains of low- and high-density phases occurs even in active
systems with only repulsive interactions among the active particles.
The Langevin and Fokker-Planck models for such active Brownian systems include
terms to account for particle propulsion in a given direction, forces arising
from direct repulsive interactions, and thermal noise.
Hydrodynamic effects are not considered.
The Langevin equations take the form
\begin{eqnarray}\label{eq:ABM}
  \frac{d}{dt} {\bf R}_i&=& \mu ({\bf F}^\text{p}_i +{\bf F}_i) +{\bf f}^\text{t}_i , \nonumber \\
  \frac{d}{dt} \hat{\bf z}_i &=& {\bf f}^\text{r}_i \times \hat{\bf z}_i
\end{eqnarray}
Here ${\bf R}_i$ and $\hat{\bf z}_i$ are the position and orientation of
motor $i$, the propulsion force is ${\bf F}^\text{p}_i = \zeta V_z \hat{\bf z}_i$,
and the mobility is $\mu = 1/\zeta$, with $\zeta$ being the friction coefficient.
The random functions ${\bf f}^\text{t}_i$ and ${\bf f}^\text{r}_i$ satisfy
fluctuation-dissipation relations,
$\langle {\bf f}^\text{t}_i(t) {\bf f}^\text{t}_j(t') \rangle = 2 D_0 \delta_{ij} {\bf 1}\delta(t-t')$ and
$\langle {\bf f}^\text{r}_i(t) {\bf f}^\text{r}_j(t') \rangle = 2 D_0^\text{R} \delta_{ij} {\bf 1}\delta(t-t')$,
where $D_0 = k_\text{B}T/\zeta$ and $2D_0^\text{R} = 1/\tau_\text{R}$ are the
translational and rotational diffusion coefficients, respectively, for the
inactive motor.

Numerical and analytical studies of such models have shown that phase
separation is favored when the speed $|V_z|$ is a decreasing function of the
active Brownian particle density.%
~\cite{fily2012,redner2013,cates2013,stenhammar2013,stenhammar2014,bialke2013,wysocki2014,fily2014,speck2014,speck2015}
Important parameters that control whether the system will segregate into
distinct phases are the P{\'e}clet number,
$\text{Pe} = |V_z| \tau_\text{R}/R_\text{m}$,
and the volume or area fraction, $\phi$.
(We consider two-dimensional systems where $\hat{\bf z}_i =
(\cos\theta_i,\sin\theta_i)$.)
Phase diagrams have been constructed that show regions in the Pe--$\phi$
parameter plane where segregation into dense and gas-like phases occurs.%
~\cite{speck2014,stenhammar2014,speck2015}
The effective diffusion coefficient for the active Brownian particles is
predicted to depend on the area fraction as $D_\text{m}(\phi) =
D_0 + \frac{1}{2} V_z(\phi)^2 \tau_\text{R}$.
This form is consistent with active Brownian model simulation results, which
also show that $|V_z(\phi)|$ decreases linearly with $\phi$.%
~\cite{stenhammar2014}

Two-dimensional simulations of a Brownian dynamics model for symmetric
sphere-dimer motors that includes reorientation effects due to conservative
forces have also been carried out.~\cite{gonnella2014}
This model is very similar to that in Eq.~\eqref{eq:ABM}, except applied to the
monomers in the dimer, and its overdamped limit takes the form
\begin{equation}
  \frac{d}{dt} {\bf R}_{i \nu}= \mu ({\bf F}^\text{p}_{i\nu} +{\bf F}_{i \nu})+{\bf f}^\text{t}_{i \nu},
\end{equation}
where $\nu$ labels the monomers in the dimer.
The propulsion force ${\bf F}^\text{p}_{i\nu}$ again acts along the dimer bond.
Phase separation dynamics is observed for certain values of Pe and $\phi$.

Some features of the collective motion of the backward-moving sphere-dimer
motors considered in this paper are captured by such Brownian dynamics models.
In particular, the dependence of the effective diffusion coefficient on area
fraction is qualitatively similar to that of active Brownian particles although
there are differences.
The motor speed is a decreasing function of $\phi$ but there are deviations
from the linear behavior seen in active Brownian models.
The main difference in the effective diffusion coefficient arises from the
dependence of $\tau_\text{R}$ on $\phi$, in contrast to the active Brownian
models which assume that $\tau_\text{R}$ is independent of $\phi$.
Taking this effect into account, $D_\text{m}(\phi)$ of the sphere-dimer can
be written again as a simple density-dependent generalization of that for a
single motor,
\begin{equation}\label{eq:Dm}
  D_\text{m}(\phi) = D_0(\phi) + \frac{1}{2} |V_z(\phi)|^2 \tau_\text{R}(\phi).
\end{equation}
This theoretical estimate is plotted in Fig.~\ref{fig:backward_diffusion}, and
one can see that the results agree well with the simulations.
In fact, the main discrepancy occurs for $\phi = 0$ where the statistical
uncertainty is largest.

However, these simple Brownian particle models cannot fully describe the
collective dynamics of chemically powered sphere-dimer motors.
For example, for the symmetric sphere-dimer motors with beads of identical
size, the Brownian particle models cannot distinguish between forward and
backward motions, yet as Fig.~\ref{fig:symmetric_cluster} shows, forward-moving
motors yield large cluster formation, but no such cluster formation occurs for
backward-moving motors.
These differences arise from interactions due to chemical gradients which are
not included in many active Brownian particle models.

Phenomenological models, analogous to those of active Brownian particles, but
incorporating the effects of chemical concentration gradients and again
neglecting hydrodynamic interactions, have been used to describe the collective
behavior of diffusiophoretic motors.
In particular, for the $\text{A}\to\text{B}$ motor reactive dynamics considered
here, the equations describing the position and orientation of motor $i$ are
supplemented with terms involving concentration gradients,
\begin{eqnarray}
  \frac{d}{dt} {\bf R}_i &=& V_z(c_\text{B}) \hat{\bf z}_i
  -a_1 {\bm\nabla} c_\text{B}({\bf R}_i) +\mu {\bf F}_i +{\bf f}^\text{t}_i \\
  \frac{d}{dt} \hat{\bf z}_i &=&
  -a_2 (1- \hat{\bf z}_i \hat{\bf z}_i) \cdot{\bm\nabla} c_\text{B}({\bf R}_i)
  +\mu{\bf F}_i \times \hat{\bf z}_i +{\bf f}^\text{r}_i \times \hat{\bf z}_i ,\nonumber
\end{eqnarray}
where $c_\text{B}({\bf r})$ is the local concentration of the product B and the
parameters $a_{1,2}$ characterize the strength of the concentration gradient
coupling.~\cite{anderson1989,anderson1991}
The motor velocity $V_z(c_\text{B})$ is a function of the concentration.
Depending on the values of the parameters $a_{1,2} \propto \frac{k_\text{B}T}{\eta}
\Lambda$, different types of collective behavior, including gas-like, dynamic
cluster states and other time-dependent and collapsed cluster states can
exist.~\cite{pohl2014,saha2014,pohl2015}

\subsection*{Modifications of microscopic dynamics}
Given the above summary of these active Brownian models that neglect
hydrodynamic interactions, it is interesting to examine the consequences of
this neglect in the context of our microscopic model.
Here we consider two modifications of the microscopic dynamics where momentum
is no longer locally conserved so that hydrodynamic interactions are absent,
and the diffusiophoretic propulsion force that depends on concentration
gradients is replaced by a constant propulsion force.

Coupling to the hydrodynamic fluid velocity fields can be eliminated by
replacing the collision step in MPCD with a global Andersen thermostat.
Instead of multiparticle collisions, the post-collision solvent velocities are
assigned from a Boltzmann distribution at temperature $T$, which locally
destroys conservation of momentum.
The motors experience the same forces by self-produced chemical gradients as in
the full dynamics; however, in the absence of local momentum conservation,
these forces on the motor are not balanced by solvent flow and the
diffusiophoretic mechanism no longer operates.
Coupling to fluid hydrodynamic modes also changes other dynamical properties of
the motor, such as its reorientation time, diffusion coefficient, and
corresponding friction coefficient (see Table~\ref{tab:modified}).
Thus, neglect of coupling to fluid hydrodynamic modes in the microscopic theory
alters both the dynamics of a single motor and its interactions with other
motors.
This is reflected in the significantly reduced values of the single-dimer
propulsion velocities given in Table~\ref{tab:modified}, which are
approximately 30--40 times slower than their values when momentum is locally
conserved.
In phenomenological models, single motor properties are effectively rescaled
while neglecting hydrodynamic coupling among motors.

Within the context of this microscopic model without hydrodynamic interactions,
one may explore the effects of chemical coupling on the collective dynamics.
Effects due to chemical gradients can be eliminated by removing the reactive
collisions that produce product in the $\text{A}\to\text{B}$ reaction and
replacing them with an external, constant force $F_\text{m}$ applied along
the bond of each sphere dimer.
The magnitude of the force is chosen such that a single dimer at area fraction
$\phi\to 0$ has the same propulsion velocity as a single dimer propelled by a
self-produced chemical gradient.
The addition of external forces implies viscous heating of the solvent, which
may be avoided by using a global Anderson thermostat that also eliminates
hydrodynamic interactions.

\begin{figure}[htbp]
  \centering
  \includegraphics{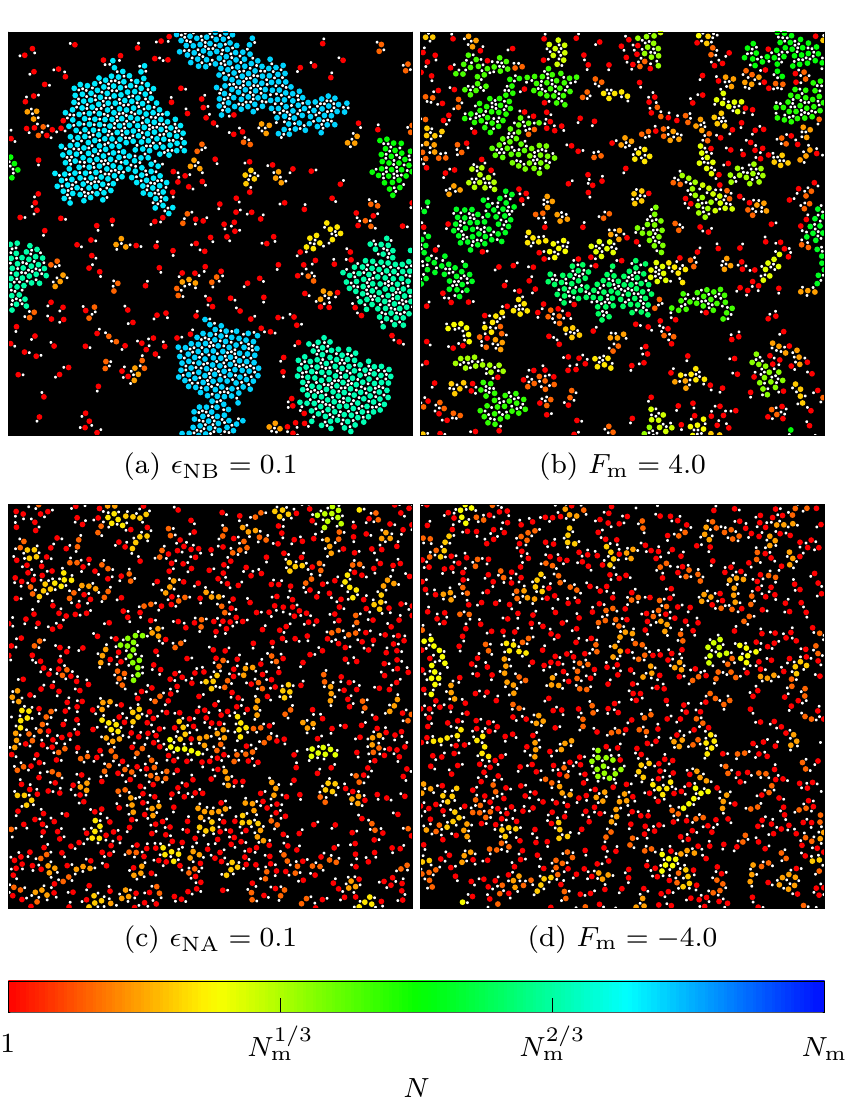}%
  \subfloat{\label{fig:modified_cluster_gradient_forward}}%
  \subfloat{\label{fig:modified_cluster_force_forward}}%
  \subfloat{\label{fig:modified_cluster_gradient_backward}}%
  \subfloat{\label{fig:modified_cluster_force_backward}}%
  \caption{Instantaneous configurations of 1000 forward-moving (top) and
    backward-moving (bottom) motors with diameters $d_\text{C} = 4$ and
    $d_\text{N} = 8$ for area fraction $\phi = 0.2$ and two modified dynamical
    models: with chemical gradients and without hydrodynamics (left), and with
    external force and without hydrodynamics (right).
    The results are shown at time $t = \num{2e6}$.
    Clusters of $N$ motors are colored to a logarithmic scale.}
  \label{fig:modified_cluster}
\end{figure}

Fig.~\ref{fig:modified_cluster} compares simulation results using the two
modified dynamical schemes described above for asymmetric sphere-dimer motors
with $d_\text{C} = 4$ and $d_\text{N} = 8$.
When chemical gradients are taken into account but hydrodynamic interactions
are neglected, for forward-moving motors, one observes the formation of large
clusters that persist for long times
(Fig.~\ref{fig:modified_cluster_gradient_forward}).
Similarly, for backward-moving motors, a statistically stationary regime
involving small transient clusters is observed
(Fig.~\ref{fig:modified_cluster_gradient_backward}).
These behaviors are qualitatively similar to those observed in the
phenomenological Langevin models that account for chemical gradients but
neglect hydrodynamic interactions; however, the time scales of the dynamics are
different since neglect of hydrodynamic interactions at the microscopic level
destroys the diffusiophoretic mechanism as discussed earlier.

When replacing chemical coupling by an applied constant force for propulsion,
in addition to neglect of hydrodynamic interaction, for forward-moving motors
instead of the formation of large clusters one observes transient clusters with
a significant amount of translational and rotational motion
(Fig.~\ref{fig:modified_cluster_force_forward}).
The collective behavior of backward-moving motors is qualitatively similar to
that of the full microscopic model
(Fig.~\ref{fig:modified_cluster_force_backward}).

Recall that for symmetric sphere-dimer motors, there is no difference between
forward- and backward-moving motors when chemical gradients are neglected.
Thus, the difference in the results for forward-
(Fig.~\ref{fig:modified_cluster_force_forward}) and backward-moving
(Fig.~\ref{fig:modified_cluster_force_backward}) asymmetric motors obtained
using an applied force to move the motor instead of a chemical gradient is a
geometric effect arising from the motor asymmetry.
For full microscopic dynamics or models that include chemical gradients but no
hydrodynamics, chemical gradients play a major role in determining the
dynamical structure, but geometrical effects arising from direct intermolecular
interactions may also be important.
The relative importance of these two effects merits further study.

\section{Conclusion}\label{sec:conc}
Through the study of microscopic models of active systems that include both the
active particles and solvent species, insight into the various factors that
determine the character of the collective behavior can be obtained.
For the diffusiophoretic sphere-dimer motors considered in this paper, these
factors include the explicit microscopic description of the reaction kinetics
on the motor surface, direct motor--motor anisotropic interactions, hydrodynamic
coupling arising from solvent fluid flows, and thermal noise.
Of course, there is a price to be paid for this level of description: there are
limitations on the size of the system that may be studied conveniently.
Although the simulation results presented in this paper are for very large
systems comprising thousands of motors and up to $10^8$ solvent molecules, the
number of motors is several orders of magnitude smaller than the largest active
Brownian particle simulations.
As a consequence, our study was restricted to a limited number of points in
parameter space, while Brownian dynamics simulations have been used to
construct phase diagrams in large parameter regions and characterize the
scaling behavior of the phase segregation process.
Since our dynamics follows directly from the intermolecular interactions, it
can be used to test the validity of the assumptions that underlie the
phenomenological models and aid in the construction of more accurate reduced
descriptions.
The research described in this paper can be extended to other motors, such as
simpler Janus motors, to investigate further aspects of the collective
dynamics.~\cite{huang2016}

The phase behavior of active systems using Brownian dynamics models is usually
studied in parameter spaces that include quantities such as the motor velocity,
P{\'e}clet numbers, volume fraction, and diffusiophoretic coupling coefficients.
In microscopic models, these parameters are functions of the intermolecular
interactions that define the system.
For example, changes in the solvent dynamics through the MPCD collision rule in
our model will change the solvent viscosity and diffusion transport properties
and will also simultaneously change the motor velocity and reorientation time,
as well as the diffusion-influenced reaction kinetics on the motor.
Changes in the intermolecular interactions between the reactive species and the
motor (described by the $\Lambda$ factor in the continuum theories) will
simultaneously change the self-propulsion properties of the motor and the
coupling through the concentration gradients due to other motors.
If hydrodynamic interactions are suppressed in the microscopic model by
altering the collision dynamics, the basic diffusiophoretic mechanism that
causes a single motor to be self-propelled is also turned off.
Rescaling of transport properties and other alterations to the dynamics can
restore some of the effects but at the expense of using a dynamical description
that is less firmly grounded.

The other factor that plays a role in determining the character of the
collective dynamics is the manner in which the system is driven from
equilibrium.
Fuel must be supplied to the motors and product must be removed to maintain the
system out of equilibrium, and these species may be introduced at the
boundaries or globally through bulk nonequilibrium reactions as in this study.
If species are supplied or removed at the boundaries, geometry and
dimensionality will play important roles; for example, correlations arising
from many-body concentration fields can lead to non-analytic dependence of
reaction rates on the motor volume fraction.%
~\cite{tucci2004,lebenhaft1979}

Active systems present challenges because they operate out of equilibrium and
display diverse phenomena.
Theoretical models at different levels of description can be used to unravel
and understand the origins of these phenomena.

This work was supported in part by grants from the Natural Sciences and
Engineering Research Council of Canada and Compute Canada.

\bibliography{collective_dimer}

\end{document}